\documentclass{article}
\usepackage{graphicx}
\usepackage{amsmath}
\usepackage{amsfonts}
\usepackage{amssymb}
\usepackage{multirow}

\newtheorem{theorem}{Theorem}[section]

\numberwithin{equation}{section}
\begin{document}
\newcommand{\T}{\mathbb{T}} 
\newcommand{\R}{\mathbb{R}}
\newcommand{\Q}{\mathbb{Q}}
\newcommand{\N}{\mathbb{N}} 
\newcommand{\Z}{\mathbb{Z}} 
\newcommand{\tx}[1]{\quad\mbox{#1}\quad} 
\newcommand{\icd}{$icd$\,}
\newcommand{\bt}{{\mbox{\boldmath$\beta$}}}

\title{Effective conservation of energy and momentum algorithm using   switching potentials suitable for molecular dynamics simulation of thermodynamical systems}
\author{Christopher G. Jesudason
\thanks{on leave from Chemistry Department, University of Malaya, 50603 Kuala Lumpur, Malaysia.}\\
{\normalsize Laboratory of Physics and Helsinki Institute of Physics,}\\
{\normalsize P.O.Box 1100, FIN-02015 HUT, Finland.}\\
{\normalsize Email: jesu@um.edu.my, chrysostomg@gmail.com}}
\date{\normalsize 4 April, 2007}
\maketitle
\begin{abstract} During a crossover  via a switching mechanism from one 2-body potential to another as might be applied  in  modeling  (chemical)  reactions in the vicinity of   bond formation, energy violations would occur due to finite step size which  determines the trajectory of the particles relative to the potential interactions of the unbonded state by numerical (e.g. Verlet)  integration.  This problem is overcome by an algorithm which preserves the coordinates of the system for each move, but corrects for energy discrepancies by ensuring both energy and momentum conservation in the dynamics. The algorithm is tested for  a hysteresis loop reaction model with an without  the implementation of the algorithm. The tests involve checking the rate of energy flow out of the MD simulation box; in the equilibrium state, no net rate of flows within experimental error  should be observed. The temperature and pressure  of the box should also be invariant within the range  of fluctuation of these quantities. It is demonstrated that the algorithm satisfies these criteria

\noindent{\bf AMS (MSC2000) Subject Classification.} 00A71-2, 70H05, 80A20
\end{abstract}

\section{PRELIMINARIES}\label{s1}
The  dimeric  particle  reaction  simulated may be written 
\begin{equation}\label{e18}
2\text{A}\begin{array}{*{20}c}
   {k_1 }  \\
    \rightleftarrows   \\
   {k_{ - 1} }  \\

 \end{array} \text{A}_\text{2}  
\end{equation}
where $k_1$ is the forward rate constant and $k_{-1}$ is the backward rate constant. The reaction simulation was  conducted at extremely high temperatures which are off-scale and not used in ordinary simulations of $LJ$ (Lennard-Jones) fluids where normally  \cite{hal} the reduced temperatures $T^\ast$  ranges $ \sim 0.3-1.2$, whereas here, $T^\ast \sim 8.0 -16.0 $, well above the supercritical
regime of the $LJ$ fluid  At these temperatures, the normal choices for time step increments do not obtain without also taking into account energy-momentum  conservation algorithms in regions where there are abrupt changes of gradient \cite{hal,cgj1,frenk1}. The global literature does not seem to cover such extreme  conditions of  simulation with discrete time steps using the Verlet velocity algorithm. The units used here are reduced $LJ$ ones \cite{hal}. The simulation was at density $\rho=0.70$ with $4096$ atomic particls which could react. The potentials used are as given in Fig.~(\ref{fig:1}) \, where $r_b=1.20$ for  the vicinity  where the bond of the dimer is broken and where 2 free particles emerge, and $r_f=0.85$ is the point along the hysteresis potential curve where the dimer is defined to exist for two previously free particles which collide. The reaction proceeds as follows; all particles interact with the splined $LJ$ pair potential $u_{LJ}$  except for  the dimeric pair $(i,j)$ formed from particles $i$ and $j$  which interact with a harmonic-like intermolecular potential modified by a switch $u(r)$ given by 
\begin{equation}
u(r)=u_{vib}(r)s(r)+u_{LJ}\left[  1-s(r)\right]\label{e4}
\end{equation}
  where  $u_{vib}(r)$ is the vibrational potential given by
eq.(\ref{eq:5}) below
\begin{equation}
u_{vib}(r)=u_{0}+\frac{1}{2}k(r-r_{0})^{2}               \label{eq:5} %
\end{equation}
The  switching function $s(r)$ is defined as 
\begin{equation}
s(r)=\frac{1}{1+\left(  \frac{r}{r_{sw}}\right)  ^{n}}  \label{eq:6}
\end{equation}
where
\[\left\{\begin{array} {ll}
s(r)  & \rightarrow1\text{ \ \ \ \ \ if }r < r_{sw}\\
s(r)  & \rightarrow0\text{ \ \ \ \ \ for }r > r_{sw}
\end{array}
\right. .\]
The switching function becomes effective when the distance between the
atoms  approach the value  $r_{sw}$ (see Fig.~(\ref{fig:1})\,). Some of the other parameters used in the equations that follow include:  
\newline
$u_0=-10,r_0=1.0,k\sim2446$ (exact value is determined by the other
input parameters),  $n=100,r_f=0.85,r_b=1.20,\mbox{and}\; r_{sw}=1.11$.
Particles $i$ and $j$ above also interact with all other particles not bonded to it via $u_{LJ}$ .
Full simulation details are given elsewhere \cite{cgj1}; suffice to say  the activation energy at $r_f$ is extremely high at approximately $17.5$. At $r_f$, the molecular potential is turned on where at this point there is actually a crossing of the potential curves although the gradients of the molecular and free $u_{LJ}$ potentials are "`very close"'. On the other hand, at $r_b$ , the switch forces the two curves to coalesce, but detailed examination shows that there is an energy gap of about the same magnitude as the cut-off point in a normal non-splined  $LJ$ potential ($\sim 0.04$ energy units),  meaning there is no crossing of the potentials. The current algorithm is applied for both these  cross-over regions with their different mechanisms of cross-over.  The MD cell is rectangular, with unit distance along the axis ( $x$ direction) of  the cell length, whereas the breadth and height was both $1/16$, implying a thin pencil-like system where the thermostats were placed at the ends of the MD cell, and the energy supplied per unit time step $\delta t$  at both ends of the cell  (orthogonal to the $x$ axis)  in the vicinity of $x=0$ and $x=1$  maintained at temperatures $T_h$ and $T_l$ could be monitored, where this energy per unit step time is respectively $\epsilon_h$ and $\epsilon_l$. At equilibrium, (when $T_h=T_l$), the net energy supplied within statistical error (meaning 1-3 units of the standard error of the $\epsilon$ distributions ) is zero, i.e. $\epsilon_l\approx\epsilon_h\approx0$. The cell is divided up uniformly into 64 rectangular regions along the $x$ axis and its thermodynamical properties of temperature and pressure are probed. The resulting values of the $\epsilon$'s and the relative invariance of the pressure and temperature profiles would be a measure of the accuracy of the algorithm from a thermodynamical point of view at the steady state. For systems with a large number of particles such thermodynamical criteria are appropriate. The synthetic thermostats now frequently used in conjunction with "`non-Hamiltonian"' MD \cite{frenk1} cannot be employed for this type of study, where actual energy increments are sampled. The runs were for 4 million time steps, with averages taken over 100 dumps, where each variable is sampled every 20 time steps. The final averages were over the 20-100 dump values of averaged quantities.
\setcounter{figure}{0}
\begin{figure}[htbp] 
\begin{center}
 \includegraphics[width=10cm]{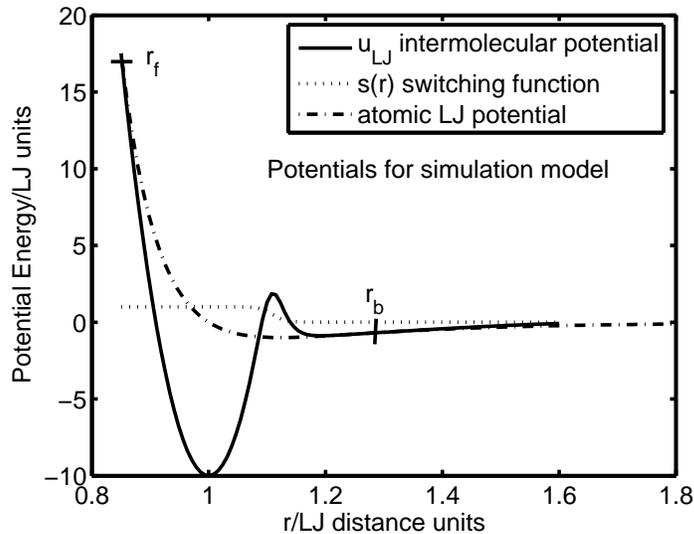}
 \caption{Potentials used for this work.}\label{fig:1}
\end{center}
\end{figure}

\begin{figure}[htbp] 
\begin{center}
 \includegraphics[width=12cm]{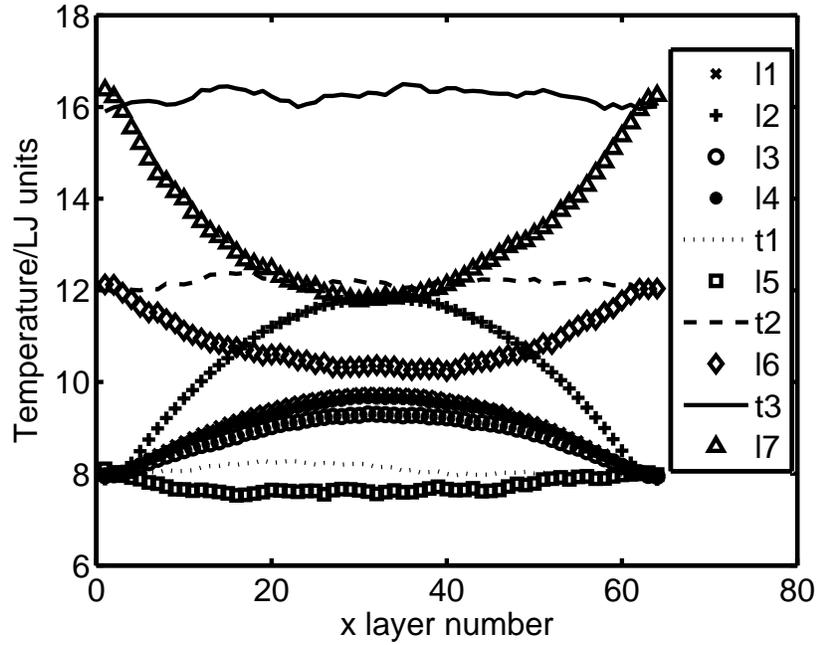}
 \caption{Temperature profile across the cell for different set conditions $a-e$ for temperature $T^\ast$ and step time  $\delta t$ pairs $(T^\ast,\delta t)$ where $a=(8.0,2.0\,ep-3),b=(8.0,5.0\,ep-4),c=(8.0,5.0\,ep-5),d=(12.0,5.0\,ep-5),e=(16.0,5.0\,ep-5)$. The curves  $\left\{\text{l}1, \text{l}3, \text{t}1, \text{t}2, \text{t}3\right\}$ results with the application of the algorithm at $r_b$ and $r_f$  with associated conditions $\text{l}1\Leftrightarrow a,\text{l}3\Leftrightarrow b, \text{t}1\Leftrightarrow c, \text{t}2\Leftrightarrow d, \text{t}3\Leftrightarrow e$  whilst the curves $\left\{l2,l4,l5,l6,l7\right\}$ are for the cases without implementing the algorithm with the associated conditions  $\text{l}2\Leftrightarrow a$,   $\text{l}4\Leftrightarrow b$, $\text{l}5\Leftrightarrow c$, $\text{l}6\Leftrightarrow d$, $\text{l}7\Leftrightarrow e$, where $ep\, x\equiv 10^x$.} \label{fig:2}
\end{center}
\end{figure}

\begin{figure}[htbp] 
\begin{center}
 \includegraphics[width=12cm]{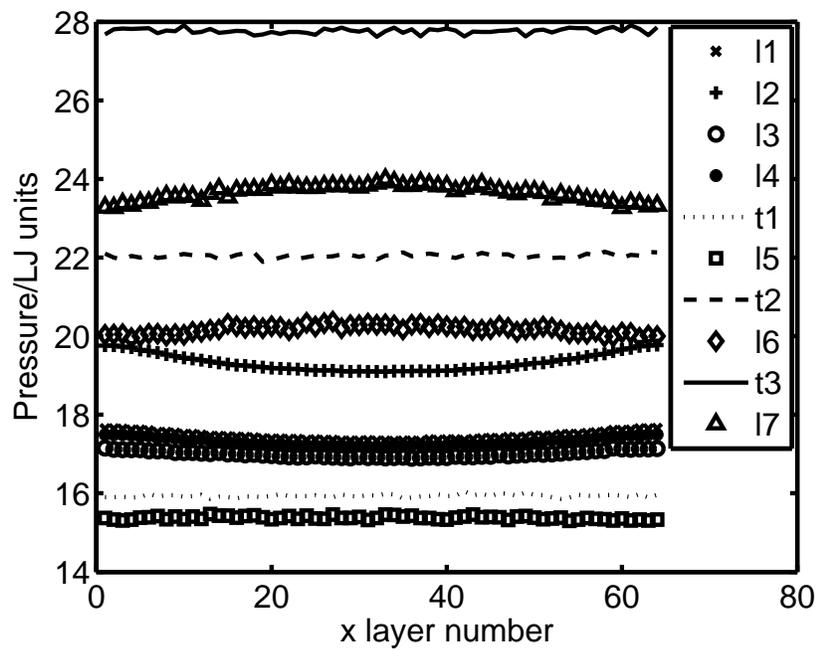}
 \caption{Pressure profile across the cell for different runs.The conditions of the runs and the labeling of the curves are exactly as in Fig.~(\ref{fig:2}).}\label{fig:3}
\end{center}
\end{figure}

\begin{table}[!h]
  \begin{tabular}{| c || c  c c |}
\hline
 Curve & $\epsilon_h$ & $\epsilon_l$ & Mean Temperature   \\
    \hline
 l1 & -.2274E+00 $\pm$0.19E-02 &-.2295E+00 $\pm$0.21E-02 &0.9063E+01 $\pm$0.62E-02    \\
   \hline
 l2 &-.5602E+00 $\pm$0.22E-02  &-.5596E+00 $\pm$0.22E-02   &0.1032E+02 $\pm$0.63E-02   \\
       \hline
 l3 &-.4161E-01 $\pm$0.14E-02   &-.4089E-01 $\pm$0.14E-02   &0.8774E+01 $\pm$0.79E-02   \\
       \hline
  l4 &-.5201E-01 $\pm$0.16E-02   &-.5103E-01 $\pm$0.17E-02  &0.8980E+01 $\pm$0.98E-02    \\
     \hline
t1  &-.5312E-03 $\pm$0.92E-03   &  -.3334E-03 $\pm$0.76E-03  &0.8082E+01 $\pm$0.49E-02   \\
    \hline
l5  &0.1311E-02 $\pm$0.82E-03  &0.1147E-02 $\pm$0.84E-03   &0.7731E+01 $\pm$0.97E-02   \\
       \hline
t2  &-.6823E-03 $\pm$0.12E-02  &-.1507E-02 $\pm$0.13E-02  &0.1216E+02 $\pm$0.17E-01  \\
       \hline
l6 &0.7291E-02 $\pm$0.12E-02  &0.6343E-02 $\pm$0.14E-02  &0.1088E+02 $\pm$0.15E-01   \\
       \hline
t3  & -.9348E-03 $\pm$0.18E-02  &-.3379E-02 $\pm$0.17E-02   & 0.1622E+02 $\pm$0.18E-01  \\ 
       \hline
l7  &0.1918E-01 $\pm$0.14E-02 & 0.1938E-01 $\pm$0.16E-02  &0.1329E+02 $\pm$0.20E-01   \\
    \hline
    \end{tabular}
\newline
\newline
  \caption{Table with values for the mean heat supply per unit step and temperature. The error is  one unit of  standard error for the quantities.  }\label{tab:1}
  \end{table}
  
The temperature $T$  and pressure $p$ are computed  by  the equipartition  and Virial expression given respectively by 
	\[ \left\langle \sum_{i=1}^N \mathbf{p_i.p_i}/m_i\right\rangle = 3Nk_B T\,\text{and}\,P =
	\rho k_BT + W/V
\]
 where $W=-\frac{1}{3}\sum_i\sum_{j>i}w(r_{ij})$ and the  intermolecular  pair Virial $w(r)$
 is given by $w(r)=r\frac{dv(r)}{dr}$ with $v$ being the potential.
  \section{ALGORITHM AND AND ANALYSIS OF NUMERICAL RESULTS}\label{s2}
  The velocity Verlet algorithm \cite[p. 81]{allen1}used here \cite{hal} and allied types generate a trajectory at time $n\delta t$ from that at $(n-1)\delta t$ with step increment $\delta t$ through a mapping $T_m$ where $(v(n\delta t), r(n\delta t))= T_m (v((n-1)\delta t), r((n-1)\delta t)  ) $ which does not scale linearly with $\delta t$. For a Hamiltonian $H$  whose potential $V$ is dependent only on  position $r$  having momentum components $p_i$, the system without external perturbation has constant energy $E$, and the normal assumption  in MD (NAMD)is that for the $n^{th}$ step,  $ \Delta E_n=\left|H(n\delta t) -E  \right|\leq \epsilon $ and also $\sum_{i=1}^N \Delta E_i \leq \epsilon^s$ for the specified $\epsilon 's$. In the simulation under NAMD, the force fields are constant and do not change for any one time step. In these cases, the energy is a constant of the motion for any time interval $\delta t_T$ when no external perturbations (e.g. due to thermostat interference) are impressed. When there is a crossing of potentials at such a time interval  interval from $\phi_b$ to $\phi_a$ at an inter particle distance(\icd)  $r_c$ (such as points $r_f$ and $r_b$ of Fig.~(\ref{fig:1})) of general particle $1$ and $2$ (the  $(1,2)$ particle pair) due to a reactive process (such as  occurs in either direction of  (\ref{e18})) a bifurcation occurs where the MD program computes the next step coordinates   as for the unreacted system (potential $\phi_b$), which needs to be corrected.  Let the \icd at time step $i$      be $r_i$ (with $\phi_b$ potential) and at step $i+1$ after interval $\delta t$ be $r_f=r_{i+1}$ where $r_f<r_c<r_i$. Due to this crossover,  a different Hamiltonian $H'$ is operative after point $r_c$ is crossed, where under NAMD, the other coordinates not undergoing crossover are not affected. For what follows, subscripts refer to the particle concerned. Let the interparticle potential at $r_f$ be $E_a=E_f=\phi_a(r_f)$ and at $r_f$ be $E_b=\phi_b(r_f)$, where $\Delta =E_b-E_a$.  Then if $r_f$ be the final coordinate due to the $\phi_b$ potential and force field, two questions may be asked: (i) Can the velocities of (1,2) be scaled, so that there is no energy or momentum violation during the crossover  based on the $\phi_b$ trajectory  calculation and (ii) Can a pseudo stochastic potential be imposed  from coordinates $r_c$ (at virtual time $t_c$) to $r_f$ such that (i) above is true? For (ii) we have
  \begin{theorem}\label{t1}
  A virtual potential which scales velocities to preserve momentum and energy can be constructed about region $r_c.$
  \end{theorem}
  \textbf{proof} 
  The external work done $\delta W$ on particles 1 and 2 over the time step is proportional to  the distance traveled since these forces are constant and  so for each of these particles $i$, $\mathbf{F_{ext,i}.\Delta r_i}=\delta W_i$ where $\mathbf{\Delta r_i}$  is the distance increment during at least part of the time step from $r_c$ to $r_f$. For the non-reacting trajectory over time $\lambda \delta t$ ($\lambda \leq 1)$ (virtual because it is not the correct path due to the crossover at $r_c$), 
\begin{equation}\label{e8}
	\delta W_2 + \delta W_1 -( \phi_b(r_f)-\phi_b(r_c))=\Delta \sum (K.E.)
\end{equation}
  where $\Delta \sum (K.E.)$ is the change of kinetic energy for the $(1,2)$ pair from the First Law between the end points $r_f,\,r_c$. Now over time interval $t_c$ to $t_f$, for the reactive trajectory, we introduce a "`virtual potential"' $V^{vir}$  that will lead to the same positional coordinates for  the pair at the end of the time step with different velocities than for the non-reactive transition leading to the transition
 \begin{equation}\label{e9}
	\delta W_2 + \delta W_1 -( V^{vir}(r_f)- V^{vir}(r_c))=\Delta \sum \,' (K.E.)
\end{equation} 
where   $\Delta \sum \,' (K.E.)$ is the change of kinetic energy for the pair with $V^{vir}$ turned on and along this trajectory, the change of potential for $V^{vir}$ is equated to the change in the K.E. of the pair as given in   the results of theorem (\ref{t2}) for all three orthogonal coordinates, i.e. 
\[\delta V^{vir}(r)- \delta \phi_b(r) =\delta\left(\Delta \sum  (K.E._{x,y,z}) -\Delta \sum \, '(K.E.)_{x,y,z} \right)\] with momentum conservation, that is $\delta V^{vir}(r_i)=\delta \phi_a(r_i)$ for the variation along the $r_i$ coordinate, but $\delta \phi_a(r_i)=-\delta K.E. $ along internuclear coordinate $ r_i $ whereas  $\delta V^{vir}=-K.E.$ (scaled about all three axes). 
 Continuity of potential implies
\begin{equation}\label{e10}
	\phi_a(r_f)=V^{vir}(r_f); \phi_a(r_c)=V^{vir}(r_c);\phi_b(r_c)=V^{vir}(r_c);
\end{equation}
Subtracting (\ref{e8}) from (\ref{e9})  and applying b.c.'s (\ref{e10}) leads to 
\begin{align}
	\Delta = \phi_b(r_f) -V^{vir}(r_f)&=&\phi_b(r_f)-\phi_a(r_f)=E_b-E_a \\ \label{e11}
	    &=& \Delta \sum \,' (K.E.) -\Delta \sum (K.E.)
\end{align}
The above shows that a conservative virtual potential could be said to be operating in the  vicinity of the transition (from $t_c \,\text{to}\, t_a $) .$\bullet$
 
   Question (i) above leads to:
   \begin{theorem}\label{t2} 
   Relative to the velocities at any $r_f$ due to the $\phi_b$ potential, the rescaled velocities $\mathbf{v\,'}$ due to the potential difference $\Delta$ leading to these final velocities due to the virtual potential can have  a form  given by    
\begin{equation}\label{e10}
 \mathbf{v_i\,'}=(1+\alpha)\mathbf{v_i}+\bt 	
\end{equation}
 (where $i=1,2$)  for a vector $\bt$.
\end{theorem}
\textbf{proof} 
From the $\mathbf{v}$ velocities at $r_f$ due to $\phi_b$ we compute the $\mathbf{v'}$ velocities at $r_f$ due to the virtual potential. Since net change of momentum is due to the external forces only, which is invariant for the $(1,2)$ pair, conservation of total momentum  relating $\mathbf{v'} $ and $\mathbf{v}$ in (\ref{e10})   yields a definition of $\bt$ ( summation from 1 to 2 for what follows, where the  mass of particle $i$ is $m_i$)
\begin{equation}\label{e11}
	{\bt}=-\alpha \sum m_i\mathbf{v_i}/\sum m_i
\end{equation}
Defining for any vector $\mathbf{s}^2=\mathbf{s.s},\bt^2=\alpha^2Q$, where 
\begin{equation}\label{e12}
 	Q=\left(\sum m_i\mathbf{v_i}\right)^2/\left(\sum m_i\right)^2
\end{equation}
then the rescaled velocities become from (\ref{e10})
\begin{equation}\label{e13}
\mathbf{v'_i}^2	=(1+\alpha)^2\mathbf{v_i}^2 + 2(1+\alpha)\mathbf{v_i}.\bt +\bt^2.
\end{equation}
With $\Delta=E_b-E_a$, Energy conservation implies
\begin{equation}\label{e14}
	\sum \frac{1}{2}m_i\mathbf{v'_i}^2-	\sum \frac{1}{2}m_i\mathbf{v_i}^2=\Delta
\end{equation}
The coupling of (\ref{e13}-\ref{e14}) leads, after several steps of algebra to 
\begin{align} \label{e15}
	\Delta =\frac{\alpha^2 m_1 m_2}{2(m_1+m_2)}\left[\mathbf{v_1}^2 + \mathbf{v_2}^2-2\mathbf{v_1}.\mathbf{v_2}\right]\\
	+\frac{2\alpha m_1 m_2}{2(m_1+m_2)}\left[\mathbf{v_1}^2 + \nonumber \mathbf{v_2}^2-2\mathbf{v_1}.\mathbf{v_2}\right] .
\end{align}
Defining $a=(\mathbf{v_1} - \mathbf{v_2})^2$,\,\, $q=m_2m_1/[{2(m_1+m_2)}]$,\,\,($q>0,a\geq0)$, then the above  is equivalent to the quadratic equation
\begin{equation}\label{e15}
	\alpha^2qa+2qa\alpha -\Delta = 0
\end{equation}
and in simulations, only $\alpha$ is unknown  and  can be determined from (\ref{e15})  where real solutions exist for $\Delta/qa\geq-1$.	$\bullet$
The above Inequality leads to a certain asymmetry concerning forward and backward reactions, even for reversible reactions where the region of formation and breakdown of molecules are located in the same region with the reversal of the sign of approximate $\Delta$. For this simulation, a reaction in either direction (formation or breakdown of dimer )  proceeds if (\ref{e16}) is true; if not then the trajectory follows the one for  the initial trajectory without any reaction (i.e. no potential crossover).
\newline 
\textbf{Interpretation of results.\,\,}Fig.~(\ref{fig:1}) shows a rapidly changing potential curve with several inflexion points used in the simulation at very high temperature (as far as I know such ranges have  not been reported in the literature for non-synthetic methods) warranting smaller time steps; larger ones would introduce errors due to the rapidly changing potential and high K.E.; thus, even with the application of the algorithm between cordinates  $r_f$ and $r_b$, curves l1 and l2 have too large a $\delta t$ value to achieve equilibrium - meaning flat or invariant - temperature (see Fig.~(\ref{fig:2}) ) or pressure (see Fig.~(\ref{fig:3}))or unit step thermostat heat supply (see Table ~\ref{tab:1})($\epsilon_h$ and $\epsilon_l$) profiles where for these curves, the $(\epsilon_h,\epsilon_l)$ values show net heat absorption; the curve at t1 (with $\delta t= 5.0\, ep -5$ show flat profiles (within statistical fluctuations and   2 standard  errors of variation) for temperature, pressure and net zero heat supply; and this choice of time step interval was found adequate for runs at  much higher temperatures ($T=12$ and $T=16$) which was used to determine thermodynamical properties \cite{cgj1}. For this $\delta t$ value and all others, no reasonable stationary equilibrium conditions could be obtained without the application of the algorithm (curves l2,l4,l5,l6 and l7). The algorithm is seen to be effective over a wide temperature range for this complex dimer reaction simulated under extreme values of thermodynamical variables and the  results here do not vary for longer runs and greater sampling statistics (e.g. 6 or 10 million time steps). The thin, pencil-like geometry of the rectangular cell with thermostats located at the ends would highlight the energy non-conservation leading to a non-flat temperature distribution, as observed and which was used to determine the regime of validity of the algorithm.

\end{document}